\documentclass[12pt]{article}

\usepackage{amsfonts}
\usepackage{hyperref}

\let\citeyear=\cite

\let\void=\unlhd

\let\l\langle
\let\r\rangle
\def\ie{i.e.}
\def\wrt{w.r.t.}
\def\var{V\!ar}
\def\qed{\hfill{\boxit{}}
  \ifdim\lastskip<\medskipamount \removelastskip\penalty55\medskip\fi}
\long\def\boxit#1{\vbox{\hrule\hbox{\vrule\kern3pt
                  \vbox{\kern3pt#1\kern3pt}\kern3pt\vrule}\hrule}}

\def\np{{\rm NP}}
\def\conp{{\rm coNP}}
\def\Dp{{\rm DP}}
\def\Dlog#1{\mbox{$\Delta^p_{#1}[\log n]$}}
\def\S#1{\mbox{$\Sigma^p_{#1}$}}
\def\P#1{\mbox{$\Pi^p_{#1}$}}

\def\nucP#1{\nuc{$\Pi^p_{#1}$}}

\def\nuc#1{\mbox{$\parallel\!\leadsto$#1}}

\def\nucDp{\nuc\Dp}

\newtheorem{lemma}{Lemma}
\newtheorem{theorem}{Theorem}

\newtheorem{definition}{Definition}

\def\proof{\noindent{\em Proof.} }

\expandafter\ifx\csname citeyear\endcsname\relax
\let\citeyear=\cite
\fi

\begin{document}

\title{On the Complexity of Second-Best Abductive Explanations}
\author{Paolo Liberatore%
\thanks{DIAG - Sapienza University of Rome,
Via Ariosto 25, 00185 Rome,
email: {\tt liberato@dis.uniroma1.it},
phone: {\tt +39 347 6906915},
corresponding author.}
\and
Marco Schaerf%
\thanks{
DIAG - Sapienza University of Rome,
Via Ariosto 25, 00185 Rome,
email: {\tt marco.schaerf@uniroma1.it}}
}

\maketitle

\begin{abstract}

When we look for abductive explanations of a given set of manifestations, an
ordering between possible solutions is often assumed. While the complexity of
optimal solutions is already known, in this paper we consider second-best
solutions with respect to different orderings, and different definitions of
what a second-best solution is.

\end{abstract}


 %

\sloppy
\section{Introduction}

The three basic reasoning mechanisms used in computational logic are deduction,
induction, and abduction~\cite{peir-55}. Deduction is the process of drawing
conclusions from information and assumptions representing our knowledge of the
world, so that the fact ``battery is down'' together with the rule ``if the
battery is down, the car will not start'' allows concluding ``car will not
start''. Induction, on the other hand, derives rules from the facts: from the
fact that the battery is down and that the car is not starting up, we may
conclude the rule relating these two facts. Abduction is the inverse of
deduction (to some extent \cite{cial-96}): from the fact that the car is not
starting up, we conclude that the battery is down. In a more complete
formalization of this environment there are many explanations for a car not
starting up. This is an important difference between abduction and deduction,
making the former, in general, more computationally complex.


A given problem of abduction may have one, none, or even many possible
solutions (explanations). Moreover, we need to perform both a consistency check
and an inference just to verify an explanation. These facts intuitively explain
why abduction is to be expected to be computationally harder than deduction.
This observation has indeed been confirmed by theoretical results. Selman and
Levesque~\citeyear{selm-leve-90,selm-leve-96} and
Bylander~et~al.~\citeyear{byla-etal-89,byla-etal-91} proved the first results
about fragments of abductive reasoning, Eiter and
Gottlob~\citeyear{eite-gott-95-a} presented an extensive analysis, Creignou and
Zanuttini~\citeyear{crei-zanu-06} and Creignou, Schmidt, and
Thomas~\citeyear{crei-schm-thom-12} classified the complexity under two kinds
of restrictions, Nordh and Zanuttini~\cite{nord-zanu-08} located the
tractability/intractability frontier, Eiter and
Makino~\citeyear{eite-maki-02,eite-maki-03,eite-maki-03-b} studied the
complexity of computing all abductive explanations, Hermann and
Pichler~\citeyear{herm-pich-10} proved the complexity of counting the number of
solutions, Fellow et al.~\cite{fell-etal-12} analyzed the problem from the
point of view of parametrized complexity. All these results proved that
abduction is, in general, harder than deduction. The analysis has also shown
that several problems are of interest in abduction. Not only the problem of
finding an explanation is relevant, but also the problems of checking an
explanation, or whether a hypothesis is in some, or all, of the explanations
(relevance). Some work on the complexity of abduction from non-classical
theories has also been
done~\cite{eite-gott-leon-97,eite-gott-leon-97-b,calv-etal-11}.

Abduction is also related to the ATMS \cite{dekl-86,reit-dekl-87} and to the
set of prime implicates of a propositional formula. Indeed, Levesque
\cite{leve-89} has proved that ATMS and prime implicates can be used to find
the abductive explanations of a literal from a Horn theory. As a result, ATMS
and algorithms for finding prime implicants of a formula can be seen as
algorithms that solve the problem of abduction; moreover, finding the prime
implicates can be seen as a preprocessing phase. Kernel resolution
\cite{delv-99} exploits the particular literals of the observation to drive the
clause generation process. Using this algorithm, Del Val has been able to
derive upper bounds on the number of generated clauses, and to prove that some
restricted classes of abduction problems are polynomial
\cite{delv-00,delv-00-b}.

Contrarily to deduction, abduction is driven by heuristic principles to best
explain the given observations. This means that even if the best possible
solution to a given problem is found, there is no warranty that it represents
the actual state. As an example, a light bulb may not turn on because it is
broken, but also because a complex set of circumstances caused a black out in
the whole town; while the first explanation is more likely and should therefore
be the preferred solution to the corresponding abduction problem, it may still
be wrong. Therefore, it may make sense not to stop at the first explanation, or
even at the set of all possible best explanations, but continue the search for
other, less likely solutions.

Other works studied the complexity of finding a solution for a problem of
abduction~\citeyear{selm-leve-90,selm-leve-96,byla-etal-89,byla-etal-91,eite-gott-95-a,crei-zanu-06,crei-schm-thom-12,nord-zanu-08}; this one considers
the problem of finding {\em another} solution once some other ones have been
found. The difference is that:

\begin{itemize}

\item in previous works, a problem of abduction is given and the task is to
find a solution;

\item in this article, a problem of abduction and a set of its solutions are
given, and the aim is to find another solution.

\end{itemize}

The difference is that the solution to be found has to be different from the
previous ones. Whenever an ordering of likeliness of explanations is given,
these solutions are assumed to be among the best ones, and the task is to find
another best explanation. The meaning of ``another best'' in this definition
may take two meaning: in the first one, we exclude the given solutions and
search for a best one among the remaining ones; in the second, we search for
another best solution of the original (unrestricted) problem. A third question
arises from the assumption that the search for the known solutions has produced
some additional data that can be used while looking for another one. The
complexity under such an assumption can be established using compilability
classes~\cite{cado-etal-02} and monotonic reductions~\cite{libe-01-jacm}. These
classification frameworks concern decision problems, which have yes/no
solutions. The specific problems considered in this article are: check if a set
of hypothesis is a solution, and check if a specific hypothesis is in some
solution.

 %

\section{Definitions}

The process of abduction starts from three elements: a propositional formula
$T$ formalizing the domain of interest, a set of variables $M$ representing the
current manifestations, and another set of variables $H$ representing their
possible explanations. In this article, abduction is formally defined as
follows.

\begin{definition}
\label{problem-of-abduction}

A {\em problem of abduction} is a triple $\l H,M,T \r$, where $T$ is a
propositional formula, $M$ is a set of propositional variables called
manifestations and $H$ is a set of propositional variables called hypotheses,
with $H \cap M=\emptyset$.

\end{definition}

Intuitively, $T$ describes how the assumptions and manifestations are related.
We know that the manifestations $M$ occur, and we want their most likely
explanation, where an explanation is a set of assumptions $A \subseteq H$ that
implies $M$ and is consistent with $T$.

\begin{definition}
\label{solution}

The {\em set of solutions} or {\em explanations} of a problem of abduction $\l
H,M,T \r$ is the set of all sets of assumptions $A \subseteq H$ such that $A
\cup \{T\}$ is consistent and $A \cup \{T\} \models M$:

\[
SOL(\l H,M,T \r)=
\{ A \subseteq H ~|~
   A \cup \{T\} \mbox{ is consistent and } A \cup \{T\} \models M \}
\]

\end{definition}

It is easy to show instances having exponentially many solutions. Ideally, each
instance should have a single solution, the assumptions that have -- in the
real world -- caused the manifestations. At least, there should be a way for
eliminating solutions that are known to be less likely than other ones.

This is achieved by employing a preorder $\preceq$ over the subsets of $H$.
Given two subsets $A,A' \subseteq H$, they are related by $A \preceq A'$ if $A$
is considered more likely than $A'$. The three preorders considered in this
article are:

\begin{itemize}

\item the cardinality-based preorder: $A \leq A'$ if and only if $|A| \leq
|A'|$, where $|.|$ denotes the cardinality of a set; in other words, $A$ is
preferred if it contains fewer assumptions than $A'$;

\item the subset-based preorder: $A \subseteq A'$; a set of assumptions
contained in another one is more likely than it;

\item the void preorder: $A \void A'$ for no pair $A,A' \subseteq H$; it
captures the case of no assumption about the relative likeliness of the
candidate solutions.

\end{itemize}

Instead of considering all solutions to a problem of abduction, one may
restrict attention to the most likely ones. Since likeliness is formalized by
$\preceq$, this amounts to consider only the minimal solutions.

\begin{definition}
\label{minimal-solutions}

The set of {\em minimal solutions} of a problem of abduction $\l H,M,T \r$ with
respect to the preorder $\preceq$ is:

\[
SOL_\preceq(\l H,M,T \r)= \min( SOL(\l H,M,T \r) , \preceq )
\]

\end{definition}

In this definition, $\min(R,\preceq)$ is the set of elements of $R$ that are
minimal with respect to $\preceq$, that is, the elements $r \in R$ such that no
$r'$ exists with $r' \preceq r$ and $r \not\preceq r'$.

The void preorder makes all solutions minimal: $SOL_\void(\l H,M,T \r)
= SOL(\l H,M,T \r)$. This allows for the notational simplification of
considering only minimal solutions, where the preorder may be $\void$,
$\leq$ or $\subseteq$.

\subsection{Second-Best Solution}

In the conditions of perfect knowledge, the set of minimal solutions of a
problem of abduction would always contain a single element: the hypotheses that
actually caused the manifestations to happen. Unfortunately, such complete
information may not be available, leading to more than one minimal solution.
Once one is found, it makes sense to continue the search for other ones. This
process is formalized as follows.

\begin{definition}
\label{second-best-solution}

Given a nonempty set of minimal solutions
{} $\{A_1,\ldots,A_m\} \subseteq SOL_\preceq(\l H,M,T \r)$
of a problem of abduction, the set of second-best solutions is:

\begin{eqnarray*}
\lefteqn{NEXT\_SOL_\preceq(\l H,M,T \r, \{A_1,\ldots,A_m\})} \\
&=& \min( SOL(\l H,M,T \r) \backslash \{A_1,\ldots,A_m\}), \preceq )
\end{eqnarray*}

\end{definition}

The case of empty set of given minimal solutions $\{A_1,\ldots,A_m\}$ is
excluded from consideration because it makes second-best solutions the same as
the minimal solutions.

\subsection{Other Best Solutions}

A second-best solution may not be a minimal solution of the original problem.
For example, if $\{A_1,\ldots,A_m\}$ includes all minimal solutions, all
second-best solutions are not minimal. This is because the definition first
excludes $\{A_1,\ldots,A_m\}$ from the set of solutions, and then takes the
minimal ones among the remaining ones. If only minimal solutions are of
interest, a different definition is more appropriate: given a set of minimal
solution, an other-best solution is a minimal solution not in the set of the
given ones.

\begin{definition}
\label{other-best-solution}

Given a nonempty set of minimal solutions
{} $\{A_1,\ldots,A_m\} \subseteq SOL_\preceq(\l H,M,T \r)$
of a problem of abduction, the set of other-best solutions is:

\[
MIN\_SOL_\preceq(P, \{A_1,\ldots,A_m\}) =
SOL_\preceq(P) \backslash \{A_1,\ldots,A_m\}
\]

\end{definition}

\subsection{Use of Additional Information}

In the formulation of the two problems of second-best solutions and other-best
solutions, we assumed that some solutions are already known. Of the computation
done to find them, what is assumed known is only the final result, that is, the
solutions. This is like discarding every intermediate data, even if it could
have been useful in the subsequent search for other solutions. For instance, if
we were able to prove (during the search for the first solutions) that an
assumption $h$ is in {\em all} solutions of the problem, then the problem of
checking other solutions is simplified (\ie, if a candidate solution does not
contain $h$, it is not a solution).

In general, we may assume that the result of the initial search is composed not
only of the first solutions, but also of some polynomially sized data
structure. This is formalized as follows: given a problem of abduction $P = \l
H,M,T \r$ and a set of previous solutions $\{A_1,\ldots,A_m\}$, is there a
polynomial-sized data structure $D$, depending only on $P$ and the known
solutions, such that verifying whether $A$ is a second-best or other best
solution is easier than the same check in which $D$ is not known?

This problem cannot be solved using the standard complexity classes, because it
involves a generic polynomially sized data structure $D$. The compilability
classes~\cite{cado-etal-02,libe-01-jacm} characterize this kind of problems.
These are summarized in Section~\ref{complexity-compilability}.

\subsection{Computational Problems of Abduction}
\label{computational-problems}

There are several computational problems that are relevant for abduction, here
we list the ones considered in this article.

\begin{itemize}

\item Existence: Decide whether a problem of abduction $P = \l H,M,T \r$ admits
a (minimal) solution, that is, $SOL(\l H,M,T \r)$ is non-empty;

\item Checking: Decide whether a set of hypotheses $A$ is a minimal
explanation, that is, whether $A \in SOL_\preceq(\l H,M,T \r)$;

\item Relevance: Decide whether a hypothesis $h$ belongs to at least a minimal
solution of a problem of abduction $P = \l H,M,T \r$, that is, $\exists A \in
SOL_\preceq(\l H,M,T \r)$ such that $h \in A$;


\end{itemize}

Finding a solution can be iteratively solved using the Relevance problem: for
every $h \in H$, if it is relevant then add it to $T$, and remove it from $H$
regardless of its relevance. The set of the relevant hypotheses iteratively
found in this manner is a solution for the abduction problem. This is therefore
a Turing reduction from solution finding to relevance checking, and gives an
upper bound to the former problem.

\subsection{Computational complexity}
\label{complexity-compilability}

The complexity analysis of the problems of second-best explanation is done in
the framework of the polynomial hierarchy and many-one polynomial reductions. A
number of books on the topic exist~\cite{bove-cres-94,sips-96,aror-bara-09}.
Decision problems (problems having a yes/no answer) are partitioned in classes
of increasing complexity. In summary, the class P contains all problems having
solving algorithm that run in time polynomial in the size of their inputs. The
class \np\  is defined in a similar way with the algorithm running on a
nondeterministic Turing machine. The class \conp\  contains all problems whose
complement (the problem with reverse yes/no answer of the original problem) is
in \np. The class \Dp\  contains all problems that can be split into a
subproblem in \np\  and one in \conp, so that the answer is yes if and only if
the answers of the two subproblems are yes. The other classes of the polynomial
hierarchy considered in this article are defined in terms of oracles, which are
subroutines whose running time is neglected. In particular, the class \S{2}\
contains all problems that are in \np\  assuming the availability of an oracle
solving a subproblem in \np. The class containing all complementary problems is
\P{2}. The class of problems solvable in polynomial time with a logarithmic
number of calls to an oracle for \S{2} is \Dlog{3}.

While membership to a complexity class is established by showing an appropriate
algorithm (running on deterministic or nondeterministic machines, using oracles
or not), proving non-membership a more difficult task. Currently, even the
existence of problems in \np\  that are not in P has never been unconditionally
proved, but only under the assumption P$\not=$\np. In particular, that
assumption implies that a problem is not in P if every other problem in \np\ 
can be reduced to it via a polynomial-time reduction. Such problems are called
\np-hard. If they also belong to \np, they are \np\  complete. The same
definitions apply to \Dp\  and \P{2}. More details about complexity classes and
reductions can be found in the cited books on computational
complexity~\cite{bove-cres-94,sips-96,aror-bara-09}.

Most hardness results in this article are proved by translating a problem of
abduction to another: for example, the problem of checking a solution to that
of checking a second-best solution. The reduction involves proving that certain
solutions of the first are turned into solutions of the second. Since being a
solution is defined in terms of satisfiability and unsatisfiability, the proofs
employ modifications that do not affect these conditions:

\begin{enumerate}

\item if a set implies a formula, the formula can be added to the set;

\item a formula entailed by the rest of a set can be removed from the set;

\item if a set contains a literal $l$ and a clause containing $l$, the latter
can be removed; clauses containing the negation of $l$ can be removed this
literal; when considering the sign of a literal, a clause written $l
\rightarrow s$ is actually $\neg l \vee s$; therefore, $l$ is negated in it;

\item if a variable $b$ only occurs in formulae that are clauses, and is
negated in all of them, these can be removed; the same if $b$ only occurs
unnegated;

\item in particular, if a variable only occurs in a single clause, that clause
can be removed;

\item if a set can be partitioned in subsets not sharing variables, it is
satisfiable if and only if each of the subsets is;

\item renaming variables does not affect satisfiability: if $X$ and $X'$ are
two sets of variables in bijective correspondence and $T$ a formula, the
formula $T[X'/X]$ obtained from $T$ by replacing each variable in $X$ with its
corresponding variable in $X'$ is satisfiable if and only if $T$ is.

\end{enumerate}

Compilability classes characterize the complexity when preprocessing part of
the data is possible~\cite{cado-etal-02,libe-01-jacm}. In fact, many
computationally hard problems, such as abduction in logical knowledge bases,
are such that part of an instance is known well before the rest of it, and
remains the same for several subsequent instances of the problem. In these
cases, it might be useful to preprocess off-line (compile) this known part so
as to simplify the remaining on-line problem. Compilability classes aim at
characterizing the complexity of problems when preprocessing is allowed for
free (it does not contribute to the complexity). For example, since P is the
class of problem solved in polynomial time, the class \nuc{P} contains all
problems that can be solved in polynomial time after preprocessing part of the
data. Hardness of these classes are defined in a different way than for the
usual complexity classes. However, in many cases hardness can be established as
follows: to prove that a problem $B$, composed of a fixed part and a varying
part, is hard for some class of compilability, exhibit a problem $A$ that is
hard for the corresponding class of complexity (for example, \np\  for
\nuc\np), such that:

\begin{enumerate}

\item there exists three polynomial-time functions $Class:S \rightarrow {\bf
N}$, $Repr:{\bf N} \rightarrow S$ and $Exte:S \times {\bf N} \rightarrow S$,
where ${\bf N}$ is the set of natural numbers and $S$ the set of valid inputs
to $A$, such that $Class(s)$ is between $0$ and the size of $s \in S$,
$Class(Repr(n))=n$ for every $n \in {\bf N}$, the answer of $A$ on $Exte(s,n)$
is yes if and only if this is the case for $s$;

\item there exists a polynomial-time reduction from $A$ to $B$ such that, the
fixed part $f$ of $B$ can be replaced by $Repr(Class(f))$
without altering the solutions of $B$.

\end{enumerate}

The three functions are called classification, representative and extension
functions. The second condition is called representative equivalence. As an
example, let $B$ be the problem of deciding whether a clause $c$ is a
consequence of a propositional formula $F$ ($F \models c$), where $F$ is the
fixed part (the part that is known in advance and can be preprocessed) and $c$
is the varying part (only known online), and $A$ the problem of deciding
whether a 3CNF formula $T$ is satisfiable. In this case, we can define the
classification function $Class(T)$ as the function that returns the number of
propositional variables in $T$, $Repr(n)$ is the function that computes the
formula containing all possible distinct 3-clauses over $n$ propositional
variables. By construction, $Class(Repr(n))=n$ for every $n \in {\bf N}$. We
can define $Exte(T,n)$ as follows: let $m<n$ be the number of variables of $T$,
we introduce $k = n - m$ new variables and add to $T$ the clause $v \vee \neg
v$ for each of them. The existence of classification, representative and
extension functions together with the representative equivalence property
guarantee that it is possible to transform any instance $(f,v)$ of the problem
$B$ into one $(Repr(Class(f)),v)$ where the fixed part only depends on the size
of $f$ but is otherwise constant. This property allows us to show that, if the
problem $B$ is compilable than the problem $A$ would become polynomial.
More details would make this introduction longer than the original content of
this article. The reader is therefore referred to other articles on
compilability classes~\cite{cado-etal-02,libe-01-jacm} for more explanations
and for examples.

For both complexity and compilability, the analysis is performed by turning
search problems into decision problems: from finding a solution to verifying
it. In the case of abduction, a decision problem is to check whether a subset
of $H$ is a minimal solution; finding a solution may instead be solved by
repeated solving the problem of {\em relevance}: checking the existence of a
minimal solution containing a given $h \in H$. This and the corresponding
problem of dispensability (no minimal solution contains $h$) have been
analyzed by Eiter and Gottlob~\cite{eite-gott-95-a}. In this article, the
problem of relevance is considered with the additional assumption that some
solutions are already known, possibly with additional information attached.

 %

\section{Second-Best Solution}

In this section we consider the problem of the second-best solutions, as
formalized by Definition~\ref{second-best-solution}: given a set of minimal
solutions, find one that is minimal among the other ones. As common in
computational complexity studies, this search problem is turned into a
verification problem in order to evaluate its complexity: given an instance of
abduction, a set of solutions and a candidate solution, check whether the
latter is a second-best solution. A solution can be found by repeatedly solving
problems of relevance, which are also analyzed.

The technical means to prove the hardness of these problems is the following
lemma, showing how to introduce a new minimal solution to a problem of
abduction.

\begin{lemma}
\label{adding}

For every problem of abduction $P$ not containing variables $s$ and $r$, a
different problem $P'$ can be built in polynomial time such that:

\[
SOL(P') = \{s\} \cup \{ A \cup \{r\} ~|~ A \in SOL(P) \}
\]

\end{lemma}

\proof Let $P=\l H,M,T \r$ be the original problem of abduction not containing
the variables $s$ and $r$. The problem $P'=\l H',M',T' \r$ is defined as
follows, where $t$ is a fresh variable and $H''$ is a set of fresh variables in
bijective correspondence to $H$:

\begin{eqnarray*}
H' &=& H \cup \{r,s\}\\
M' &=& \{t\}\\
T' &=&
	(T[H''/H] \vee \neg r) \wedge
	\bigwedge \{ h \rightarrow h'' ~|~ h \in H \} \wedge
	((r \wedge \bigwedge M) \rightarrow t) \wedge			\\
&&
	(\neg s \vee t) \wedge
	(\neg s \vee \neg r) \wedge 
	\bigwedge \{\neg s \vee \neg h ~|~ h \in H \}
\end{eqnarray*}

The claim is proved in three steps: first, $s$ is a solution of $P'$; second,
every solution of $P$ is also a solution of $P'$ with the addition of $r$;
third, every solution of $P'$ is either $s$ or a solution of $P$ with $r$ added
to it.

Since $T'$ contains $\neg s \vee t$ and $\neg s \vee \neg r$, the union $\{s\}
\cup \{T'\}$ implies $t$ and $\neg r$, and can therefore be by removing all
clauses containing one of these literals, resulting in a satisfiable set:

\begin{eqnarray*}
\{s\} \cup \{T'\}
&\equiv&
	s \wedge
	\bigwedge \{ h \rightarrow h'' ~|~ h \in H \} \wedge
	t \wedge \neg r \wedge
	\bigwedge \{\neg h ~|~ h \in H \}				\\
&\equiv&
	s \wedge
	t \wedge
	\neg r \wedge \bigwedge \{\neg h ~|~ h \in H \}
\end{eqnarray*}

The second part of the proof shows that if $A \in SOL(P)$ then $A \cup \{r\}
\in SOL(P')$. Since $T'$ contains $\neg s \vee \neg r$, the union $\{r\} \cup
\{T'\}$ implies $\neg s$. All clauses containing $\neg s$ can therefore be
removed, as well as $\neg r$ from the clauses containing it:

\[
A \cup \{r\} \cup \{T'\} \equiv
	\bigwedge A \wedge r \wedge
	T[H''/H] \wedge
	\bigwedge \{ h \rightarrow h'' ~|~ h \in H \} \wedge
	\neg s \wedge
	((\bigwedge M) \rightarrow t)
\]

Since $A$ is a solution of $P$, then $A \cup \{T\}$ has a model. This model can
be extended to satisfy $A \cup \{r,T\}$ by setting each $r$ to true, $s$ to
false and $h'' \in H''$ to the same value of the corresponding $h \in H$.

Since $A \cup \{T\} \models M$ and $A \cup \{r,T'\}$ imply $h \rightarrow h''$,
$T[H''/H]$ and $(\bigwedge M) \rightarrow t$, it follows that $A \cup \{r,T'\}
\models t$. This proves that $A \cup \{r\}$ is a solution of $P'$.

The final part of the proof is to show that $P'$ has no other solution beside
$\{s\}$ and $A \cup \{r\}$ where $A$ is a solution of $P$. Since $T'$ includes
$\neg s \vee \neg r$ and $\neg s \vee \neg h$ for every $h \in H$, it follows
that $\{s\} \cup \{T'\}$ entails the negation of every variable in $H'$ but
$s$; therefore, no solution contains $s$ except $\{s\}$.

Regarding the other solutions, it is now proved that a subset $A' \subset H'$
that is satisfiable with $T'$ but contains neither $s$ nor $r$ is not a
solution. Indeed, if $s,r \not\in A'$ then these two variables only occur
negated in $A' \cup \{T'\}$, and all the clauses containing them can therefore
be removed, leading to the following formula:

\[
\bigwedge A' \wedge \bigwedge \{ h \rightarrow h'' ~|~ h \in H \}
\]

This formula does not contain $t$, therefore it does not imply it. This proves
that every solutions contain either $s$ or $r$. Since no solution contain both
variables thanks to $\neg s \vee \neg r$, a solution not containing $s$ is in
the form $A \cup \{r\}$ with $A \subseteq H$. Remains to be proved that $A$ is
a solution of $P$, in this case.

Since $T'$ contains $\neg s \vee \neg r$, it follows that $A \cup \{r,T'\}$
implies $\neg s$. Therefore, all clauses containing $\neg s$ can be removed:

\begin{eqnarray}
\nonumber
\lefteqn{\{A\} \cup \{r,T'\} \equiv}					\\
\nonumber
&\equiv&
	\bigwedge A \wedge r \wedge
	(T[H''/H] \vee \neg r) \wedge
	\bigwedge \{ h \rightarrow h'' ~|~ h \in H \} \wedge
	((r \wedge \bigwedge M) \rightarrow t)				\\
\nonumber
&\equiv&
	\bigwedge A \wedge r \wedge
	T[H''/H] \wedge
	\bigwedge \{ h \rightarrow h'' ~|~ h \in H \} \wedge
	((\bigwedge M) \rightarrow t)					\\
\nonumber
&\equiv&
	\bigwedge A \wedge r \wedge
	T[H''/H] \wedge
	\bigwedge \{ h'' ~|~ h \in A \} \wedge				\\
\label{new-variable}
&&
	\bigwedge \{ h \rightarrow h'' ~|~ h \in H \backslash A \} \wedge
	((\bigwedge M) \rightarrow t)
\end{eqnarray}

Since renaming does not affect satisfiability, variables $H$ and $H''$ can be
swapped, making $\{h'' ~|~ h \in A\}$ become $A$ and $T[H''/H]$ become $T$.
What results is a set containing $A \cup \{T\}$, which is therefore
satisfiable. This is the first condition for $A$ being a solution of $P$.

The second part is $A \cup \{T\} \models M$. Since $A \cup \{r,T'\} \models t$,
the set $A \cup \{r,T',\neg t\}$ is inconsistent. Thanks to
Equivalence~(\ref{new-variable}), it can be rewritten:

\begin{eqnarray*}
\lefteqn{A \cup \{r,T',\neg t\} \equiv}					\\
&\equiv&
	\bigwedge A \wedge r \wedge
	T[H''/H] \wedge
	\bigwedge \{ h'' ~|~ h \in A \} \wedge				\\
&&
	\bigwedge \{ h \rightarrow h'' ~|~ h \in H \backslash A \} \wedge
	((\bigwedge M) \rightarrow t)
	\wedge \neg t							\\
&\equiv&
	\bigwedge A \wedge r \wedge
	T[H''/H] \wedge
	\bigwedge \{ h'' ~|~ h \in A \} \wedge				\\
&&
	\bigwedge \{ h \rightarrow h'' ~|~ h \in H \backslash A \} \wedge
	\neg (\bigwedge M)
	\wedge \neg t							\\
\end{eqnarray*}

Formulae $\bigwedge A$, $r$, $\neg s$, $\neg t$ and $h \rightarrow h''$ with $h
\not\in A$ contain variables occurring only once in the set. Removing them
results in $T[H''/H] \wedge \neg \bigwedge M \wedge \bigwedge \{h'' ~|~ h \in
A\}$. By renaming $H''$ to $H$, this is $\bigwedge A \wedge T \wedge \neg
\bigwedge M$. Its unsatisfiability implies $A \cup \{T\} \models M$.~\qed

This lemma shows how to add the new solution $\{s\}$ to a given problem of
abduction. This addition makes the problem of finding a solution in the old
instance equivalent to finding a solution different from $\{s\}$ in the new
one. The solution $\{s\}$ is minimal with respect to the three considered
orderings, since no solution of the form $\{r\} \cup A$ is contained or has
less literals than it. Since the problem modification can be performed in
polynomial time, it shows that if the problem of checking a minimal solution is
hard for some complexity class, then the corresponding problem of second-best
solution checking is hard for the same class. As a result, in the following
complexity characterizations of the second-best solution problems the hardness
parts are all proved by a simple reference to this lemma.

This lemma provides a reduction from the problem of checking whether $H \in
SOL_\void(\l H,M,T \r)$ to that of checking whether $H \in NEXT\_SOL_\void(\l
H,M,T \r, \{A_1,\ldots,A_m\})$, therefore proving the hardness of the second
problem from the hardness of the first. Verifying a solution with the empty
preorder is mentioned to be \Dp-hard by Eiter and
Gottlob~\cite{eite-gott-95-a}, but as far it was possible to verify no formal
proof was published to date. The claim is proved for the particular candidate
solution $\emptyset$; since this is minimal if it is a solution, hardness holds
for all considered orderings.

\begin{lemma}
\label{empty-hard}

Checking whether $\emptyset \in SOL(\l H,M,T \r)$ is \Dp-hard.

\end{lemma}

\proof This property is stated by Eiter and Gottlob~\cite{eite-gott-95-a} for
an arbitrary candidate solution as an easy corollary of their results, but as
far as we know, no proof has been published, possibly because of its extreme
simplicity: by translating formulae $F$ and $G$ over variables $X$ into the
problem of abduction $\l \emptyset, \{m\}, T\r$, where $T=F \wedge (\neg
G[X'/X] \rightarrow m)$, $X'$ is a set of fresh variables in one-to-one
correspondence with $X$ and $m$ a fresh variable. This is a reduction from the
{\sl sat-unsat} problem of checking whether $F$ is satisfiable and $G$ is
unsatisfiable to the problem of checking whether $\emptyset$ is a solution of
$\l \emptyset,\{m\},T\r$. Indeed, $\emptyset \cup \{T\}$ is equivalent to $F
\wedge (\neg G[X'/X] \rightarrow m)$. This formula is satisfiable if and only
if $F$ is satisfiable, since the rest is satisfied by the model where $m$ is
true. This means that $\emptyset$ is a solution if and only if $F$ is
satisfiable and $\emptyset \cup \{T\} \models m$. The latter condition is
equivalent to the unsatisfiability of $F \wedge (\neg G[X'/X] \rightarrow m)
\wedge \neg m$, which is equivalent to $F \wedge G[X'/X] \wedge \neg m$. Since
$F$ is satisfiable and does not share variables with the rest of the formula,
and the same for $\neg m$, the formula is unsatisfiable if and only if
$G[X'/X]$ is unsatisfiable. Since satisfiability is unaffected by variable name
change, this proves that $\emptyset$ is a solution of $\l \emptyset, \{m\},T
\r$ if and only if $F$ is satisfiable and $G$ is unsatisfiable. This reduction
proves that the problem is \Dp-hard.~\qed

The complexity of checking whether a set of hypotheses is a solution is an
easy consequence of this lemma.

\begin{theorem}

Checking whether $A \in SOL(\l H,M,T \r)$ is \Dp-complete.

\end{theorem}

\proof Membership follows from the problem being defined as the satisfiability
of $A \cup \{T\}$ and the unsatisfiability of $A \cup \{T,\neg \bigwedge M\}$.
Lemma~\ref{empty-hard} proves that the problem is hard even in the particular
case $A=\emptyset$.~\qed

Together with Lemma~\ref{adding}, this result proves that the
second-best solution problem is \Dp-hard for $\void$. It is also a member of
this class, as the following theorem proves.

\begin{theorem}

Deciding whether $A \in NEXT\_SOL_\void(\l H,M,T \r, \{A_1,\ldots,A_m\})$ is
\Dp-complete.

\end{theorem}

\proof By definition, $\void$ is the empty preorder: $A \void A'$ never holds.
All solutions are minimal according to this preorder. Reworded: the set of
minimal solutions coincides with the set of all solutions.

The problem is in \Dp\  because it can be solved by first checking whether $A
\cup \{T\} \models M$ and then whether $A \cup \{T\}$ is consistent and $A$ is
different from each element of $\{A_1,\ldots,A_m\}$. The subproblem $A \cup
\{T\} \models M$ is in \conp. The rest of the problem can be solved by
nondeterministically generating every possible propositional model over the
considered variables and checking whether it satisfies $A \cup \{T\}$ and
whether $A$ is different from each element of $\{A_1,\ldots,A_m\}$; both steps
can be done in polynomial time; as a result, the problem is in \Dp.

Hardness is a consequence of Lemma~\ref{empty-hard}, since $\emptyset$ is
minimal with respect to set cardinality. As a result, $\emptyset$ is a solution
if and only if it is a $\leq$-minimal solution.~\qed

Relevance is harder than verification. Intuitively, the complexity increase is
due to the necessity of searching for a solution, among the possibly many ones,
that contains the hypothesis $h$ to be checked for relevance.

\begin{theorem}

Given $\l H,M,T \r$ and $h \in H$, deciding the existence of $A$ such that $h
\in A$ and $A \in NEXT\_SOL_\void(\l H,M,T \r, \{A_1,\ldots,A_m\})$ is
\S{2}-complete.

\end{theorem}

\proof The problem can be solved by a nondeterministic algorithm employing an
oracle for the propositional satisfiability problem. The algorithm
nondeterministically generates each possible $A \subseteq H$ and calls the
oracle for the satisfiability of $A \cup \{T\}$ and of $A \cup \{T,\neg
\bigwedge M\}$. If the first is satisfiable, the second is unsatisfiable, $h
\in A$ and $A$ is different than each element of $\{A_1,\ldots,A_m\}$, the
algorithm returns yes: $h$ is relevant. Since the nondeterministic machine
returns yes if some of its nondeterministic runs return yes, this algorithm
establishes the existence of a solution containing $h$.

Hardness is a consequence of a result by Eiter and
Gottlob~\cite[Theorem~4.1.1]{eite-gott-95-a} and Lemma~\ref{adding}. Indeed,
the lemma shows how a problem of abduction $P$ can be used to build another one
$P'$ that has the same solutions of $P$ with $\{r\}$ added to each, plus the
single new solution $\{s\}$. This provides a reduction: $h$ is in some
solutions of $P$ if and only if $h$ is in some solutions of $P'$ different from
$\{s\}$. Since the first problem is
\S{2}-hard~\cite[Theorem~4.1.1]{eite-gott-95-a}, the latter is \S{2}-hard as
well.~\qed

 %

The containment preorder $\subseteq$ limits the solutions to those that do not
include any hypothesis that could be removed, that is, the unnecessary ones.
This for example rules out $\{h_1,h_2\}$ if $\{h_1\}$ is a solution. The
additional requirement of minimality does not increase the cost of verifying a
solution, which remains \Dp-complete as for the case of the empty preorder.

\begin{theorem}

Checking whether $A \in SOL_\subseteq(\l H,M,T \r)$ is \Dp-complete.

\end{theorem}

\proof The problem is in \Dp\  because it can be solved by a number of parallel
satisfiability and unsatisfiability checks. Indeed, that $A$ is a solution is
equivalent to the satisfiability of $A \cup \{T\}$ and the unsatisfiability of
$A \cup \{T, \neg \bigwedge M\}$. The first condition implies the
satisfiability of $A' \cup \{T\}$ for every $A' \subseteq A$. As a result, $A$
is not a minimal solution only if there exists $A' \subset A$ such that $A'
\cup \{T\} \models M$. This implies $A \backslash \{h\} \cup \{T\} \models M$
for every $h \in A \backslash A'$ by monotonicity of $\models$. The converse
also holds: $A$ is not minimal if such $h$ exists, since $A \backslash \{h\}
\subset A$ for every $h \in A$. As a result, $A$ is a minimal solution if and
only if:

\begin{itemize}

\item $A \cup \{T\}$ is consistent;

\item $A \backslash \{h\} \cup \{T, \neg \bigwedge M\}$ is consistent for every
$h \in A$;

\item $A \cup \{T, \neg \bigwedge M\}$ is inconsistent.

\end{itemize}

These tests are in polynomial number and can be done in parallel by renaming
the variables. As a result, the whole problem amounts to checking whether a
formula is satisfiable and another is not.

Hardness is a direct consequence of Lemma~\ref{empty-hard}, which proves that
establishing whether $\emptyset \in SOL(\l H,M,T \r)$ is \Dp-hard. Since
$\emptyset$ is contained in every other subset of $H$, if any, it is a minimal
solution if and only if it is a solution. As a result, $\emptyset \in
SOL_\subseteq(\l H,M,T \r)$ is \Dp-hard.~\qed

Given this result, the problem of checking a second-best solution can be proved
to be complete for the same class.

\begin{theorem}
\label{setc}

Deciding whether $A \in NEXT\_SOL_\subseteq(\l H,M,T \r, \{A_1,\ldots,A_m\}\r)$
is \Dp-complete.

\end{theorem}

\proof Membership is proved as in the previous theorem, with two variants.
First, $A$ is not a second-best solution if is in $\{A_1,\ldots,A_m\}$. Second,
the check for consistency of $A \backslash \{h\} \cup \{T, \neg \bigwedge M\}$
is skipped if $A \backslash \{h\}$ is in $\{A_1,\ldots,A_m\}$.

Hardness is proved by Lemma~\ref{adding} and the previous theorem, showing the
problem with no given solution \Dp-hard. The lemma proves that $A'$ is in
$SOL(\l H',M',T' \r)$ if and only if either $A'=\{s\}$ or $A'=A \cup \{r\}$
with $A \in SOL(\l H,M,T \r)$, which means that the solution $\{s\}$ is
minimal. As a result, in $NEXT\_SOL_\subseteq(\l H',M',T' \r, \{\{s\}\})$ the
second argument $\{\{s\}\}$ is a set of minimal solutions of the first, $\l
H',M',T' \r$. The solutions of $\l H',M',T' \r$ not in $\{\{s\}\}$ are those $A
\cup \{r\}$ with $A \in SOL(\l H,M,T \r)$. Since $s$ is not in $\l H,M,T \r$, a
solution $A \cup \{r\}$ does not contain $s$, which means that it is minimal if
and only if $A$ is minimal. This is therefore a reduction from checking a
minimal solution of $\l H,M,T \r$ to that of checking a second-best solution in
$NEXT\_SOL_\subseteq(\l H',M',T' \r, \{\{s\}\})$. Since the former proved is
\Dp-hard by the previous theorem, the latter is hard for the same class.~\qed

This result establishes the complexity of verifying a solution of an abduction
problem in presence of other minimal solutions. Searching for a solution can be
turned into the decision problem of relevance (checking the existence of
solutions with a given $h \in H$) as already explained. Relevance for the
subset preorder is \S{2}-complete~\cite[Theorem 4.2.1]{eite-gott-95-a}.
Lemma~\ref{adding} shows how to carry the hardness part of this result to the
case where other minimal solutions are known.


\begin{theorem}

Existence of a solution in $NEXT\_SOL_\subseteq(\l H,M,T \r,
\{A_1,\ldots,A_m\})$ containing a given $h \in H$ is \S{2}-complete.

\end{theorem}

\proof Membership can be proved by nondeterministically generating all possible
subsets $A$ of $H$ and then checking (possibly using the oracle) whether $h \in
A$, whether $A \not\in \{A_1,\ldots,A_m\}$, whether $A \cup \{T\}$ is
consistent, whether $A \cup \{T\} \models M$ and whether $A \backslash \{h'\}
\cup \{T\} \not\models M$ for all $A \backslash \{h'\} \not\in
\{A_1,\ldots,A_m\}$ with $h' \in A$.

Hardness is a consequence of the hardness result without the given solutions
$\{A_1,\ldots,A_m\}$, since Lemma~\ref{adding} implies that $A \in
SOL_\subseteq(\l H,M,T \r)$ if and only if $A \cup \{r\} \in
NEXT\_SOL_\subseteq(\l H',M',T' \r, \{\{s\}\})$. As a result, $h$ is in some
element of $SOL_\subseteq(\l H,M,T \r)$ if and only if it is in some element of
$NEXT\_SOL_\subseteq(\l H',M',T' \r, \{\{s\}\})$. This is a reduction from
relevance without given solutions to relevance for second-best solutions,
proving the \S{2}-hardness of the latter.~\qed

 %

Let $\leq$ be the preorder of solution defined by cardinality. As for $\void$
and $\subseteq$, the hardness of the problems of verification and relevance is
proved by reducing to the them the corresponding problems without the given
solutions. The following theorem shows the complexity of the verification
problem.

\begin{theorem}
\label{cardinality-checking}

Checking whether $A \in SOL_\leq(\l H,M,T \r)$ is \P{2}-complete.

\end{theorem}

\proof Non-membership can be verified with a nondeterministic algorithm
employing an oracle for solving the satisfiability problem. Given an abduction
problem and a subset $A \subseteq H$, the algorithm nondeterministically
generates each possible $A' \subseteq H$. After this $A'$ is produced, the
following checks are done, with the help of the oracle: that either $A \cup
\{T\}$ is unsatisfiable, or $A \cup \{T, \neg \bigwedge M\}$ is satisfiable, or
the following three conditions hold: $|A'|<|A|$, $A' \cup \{T\}$ is consistent
and $A' \cup \{T, \neg \bigwedge M\}$ is inconsistent. If all these hold, then
either $A$ is not a solution or smaller solution $A'$ exists.

\

Hardness is proved by reduction from the problem of non-relevance, which Eiter
and Gottlob~\cite[Theorem~4.2.1]{eite-gott-95-a} proved to be \S{2}-complete
even if the formula $T$ is consistent~\cite[Definition~2.1.1]{eite-gott-95-a}.
Given a problem of abduction $\l H,M,T \r$ and $h \in H$, a $\leq$-minimal
solution of $\l H,M,T \r$ containing $h$ exists if and only if $S$ is not a
$\leq$-minimal solution of the problem $\l H',M',T' \r$ defined as follows.

\begin{eqnarray*}
H' &=& H \cup Z \cup S				\\
M' &=& M \cup \{w\}				\\
T' &=& T[h''/h][M''/M] \wedge
       \bigwedge \{m'' \rightarrow m ~|~ m \in M\} \wedge \\
&&
       (h \rightarrow h'') \wedge
       (h \rightarrow w) \wedge
       (\bigwedge S \rightarrow \bigwedge M')
\end{eqnarray*}

If $|H|=n$, then $S$ is a set of $n+1$ fresh variables. Also $h''$ and $w$ are
fresh variables and $M''$ is a set of fresh variables in one-to-one
correspondence with $M$.


Regardless of the original problem, $S$ is a solution of $\l H',M',T' \r$.
Indeed, $S \cup \{T'\}$ contains $S$ and $\bigwedge S \rightarrow M'$, which
imply $M'$. Remains to prove that $S \cup \{T'\}$ is consistent. By definition,
$M'=M \cup \{w\}$. All subformulae of $T'$ are entailed by $M'$ but
$T[h''/h][M''/M]$ and $h \rightarrow h''$ and can therefore be removed without
affecting consistency. Since $S$ is a set of fresh variables, none is in
$T[h''/h][M''/M] \wedge (h \rightarrow h'')$. This formula is consistent because
$T$ is consistent. As a result, $S \cup \{T'\}$ is consistent, proving that $S$
is a solution of $\l H',M',T' \r$.


The solutions of $\l H',M',T' \r$ are further characterized (as proved below)
to contain one of the following:

\begin{enumerate}

\item a solution of $\l H,M,T \r$ that contains $h$;

\item $S$.

\end{enumerate}


Since $|S|=n+1$ while a solution of the original problem has size between $0$
and $|H|=n$, it follows that $S$ is a minimal-size solution if and only if the
original problem has no solution containing $h$. This is therefore a reduction
from non-relevance to solution checking. Since relevance is \S{2}-hard, the
problem of solution checking would be \P{2}-hard. Remains to prove that every
solution of $\l H',M',T' \r$ contains one of the two sets above.



Let $A'$ be a solution of $\l H',M',T' \r$. If $A'$ does not contain $s_i \in
S$ then $s_i$ only occur negated in $A' \cup \{T'\}$ and $A' \cup \{T', \neg
\bigwedge M\}$, in particular in the formula $\bigwedge S \rightarrow \bigwedge
M$. Therefore, this formula can be removed without affecting consistency. The
other variables of $S$ may only occur once (in $A'$). They can therefore be
removed as well. This proves that if a solution of $\l H',M',T' \r$ does not
contain all of $S$ then removing all elements of $S$ from it leads to another
solution.


If $A'$ is a solution not intersecting $S$, then $A' \cap H$ is a solution of
the original problem. This is proved as follows. The set $A' \cup \{T'\}$
contains $(A' \cap H) \cup \{T[h''/h][M''/M], h \rightarrow h''\}$. Since the
first is consistent, the second is consistent as well. Replacing each $m''$
with $m$ and swapping $h$ and $h''$ transforms $T[h''/h][M''/M]$ into $T$.
Since variable name changes do not affect satisfiability, the resulting set
$(A' \cap H) \cup \{T, h'' \rightarrow h\}$ is consistent. It contains $(A'
\cap H) \cup \{T\}$, whose consistency is the first condition for $A' \cap H$
being a solution of $\l H,M,T \r$.

The second is $(A' \cap H) \cup\{T\} \models m$ for every $m \in M$. Since $A'
\cup \{T'\} \models M'$ and $M \subseteq M'$, it also holds $A' \cup \{T'\}
\models m$ for every $m \in M$. This is the same as the inconsistency of $A'
\cup \{T',\neg m\}$. Since $w$ only occurs in the clauses $h \rightarrow w$ and
$\bigwedge S \rightarrow w$, and is positive in both, these can be removed
without affecting satisfiability. The same for the variables of $S$, which only
occur negated, and the variables of $M \backslash \{m\}$, which only occur
unnegated. Some further simplifications can be done:

\begin{eqnarray*}
\lefteqn{
A' \cup \{T[h''/h][M''/M]\} \cup
\{h \rightarrow h'', m'' \rightarrow m, \neg m\}
\equiv}									\\
&\equiv&
A' \cup \{T[h''/h][M''/M], h \rightarrow h'', \neg m'', \neg m\}
\end{eqnarray*}

In this formula, $h$ and $m$ only occur once and can therefore be removed. What
remains is $A' \cup \{T[h''/h][M''/M], \neg m''\}$. Renaming $M''$ to $M$ and
$h''$ to $h$ does not affect satisfiability; therefore, the set $A' \cup \{T,
\neg m\}$ is unsatisfiable.

Since the variables in $S$ may only occur once in this set, in $A'$, they can
be removed. The result is $(A' \cap H) \cup \{T,\neg m\}$. Since the changes
did not affect satisfiability and the original set was unsatisfiable, so is
this one. As a result, $(A' \cap H) \cup \{T\} \models m$. Since this holds for
every $m \in M$, and the satisfiability of $(A' \cap H) \cup \{T\}$ was already
proved, $A' \cap H$ is a solution of $\l H,M,T \r$.

\

What remains to be proved is that either $A'$ contains $h$ or the whole $S$. To
the contrary, assume that $A'$ does not contain $h$ and does not contain some
$s_i \in S$. Since $w \in M'$, formula $A' \cup \{T'\} \wedge \neg w$ is
unsatisfiable. If $A'$ does not contain $h$ and does not contain $s_i$, these
variables occur in $A' \cup \{T'\} \wedge \neg w$ only in the clauses $h
\rightarrow w$ and $\bigwedge S \rightarrow \bigwedge M'$. All these
occurrences of $h$ and $s_i$ are negated; therefore, these clauses can be
removed without affecting satisfiability. Since these are the only subformulae
of $A' \cup \{T'\} \wedge \neg w$ containing $w$, what remains is a subformula
of $A' \cup \{T'\}$, which is consistent because $A'$ is a solution. This
contradiction proves that every solution of $\l H',M',T' \r$ contains either
$h$ or the whole $S$.~\qed

The following theorem shows the complexity of the second best solution
verification problem with the cardinality-based preorder.

\begin{theorem}

Deciding whether $A \in NEXT\_SOL_\leq(\l H,M,T \r, \{A_1,\ldots,A_m\}\r)$ is
\P{2}-complete.

\end{theorem}

\proof Membership is proved as follows: $A$ is a second-best solution if it is
in $SOL(\l H,M,R \r)$ and for every $A' \subseteq H$ such that $|A'| < |A|$ it
holds that either $A' \cup \{T\}$ is inconsistent, $A' \cup \{T\} \not\models
M$ or $A' \in \{A_1,\ldots,A_m\}$. All these checks can be done with an \np\
oracle, once a subset $A' \subseteq H$ is nondeterministically generated.

Hardness is proved by the reduction of Lemma~\ref{adding}, using $m=1$ and
$\{A_1,\ldots,A_m\}=\{\{s\}\}$. As the lemma proves, $\{s\}$ is indeed a
solution, and is also among its minimal ones because all other ones (if any)
have the form $H \cup \{r\}$, so they have cardinality larger or equal than
one.

The lemma also proves that every solution to the original problem is translated
into a solution of the new one. This reduction preserves the relative size of
explanations, as they are all added one element. As a result, the solutions are
not only all translated, but maintain their relative size. Therefore,  $A \cup
\{r\} \in NEXT\_SOL_\leq(\l H',M',T' \r, \{\{s\}\})$ holds if and only if $A
\in SOL_\leq(\l H,M,T \r)$ holds.~\qed

The problem of existence of a second-best solution with a given element of $H$
can be shown to be \Dlog{3}-complete.

\begin{theorem}

Existence of a solution in $NEXT\_SOL_\leq(\l H,M,T \r, \{A_1,\ldots,A_m\})$
containing a given $h \in H$ is \Dlog{3}-complete.

\end{theorem}

\proof The problem of checking for the existence of a solution $A$ with size
bounded by a number $k$ and containing $h$ is in \S{2}, as it amounts to
nondeterministically generating a solution and then checking it for being a
second best-solution and for its size being less than or equal to $k$. The
problem of relevance can be therefore solved by a binary search for the minimal
size of solutions~\cite[Theorem~4.3.2]{eite-gott-95-a}: start with $k=|H|/2$,
and if the result is positive change $k=|H|3/4$, otherwise $k=|H|/4$. Once the
minimal size is found, the problem can be solved by nondeterministically
generating all solutions of this size not being in $\{A_1,\ldots,A_m\}$ and
then checking whether $h$ is in some of them.

Hardness follows from Lemma~\ref{adding}: $h$ is $\leq$-relevant to $\l H,M,T
\r$ if and only if a solution of $NEXT\_SOL_\leq(\l H',M',T' \r,\{\{s\}\})$
containing $h$ exists; this is proved like in the previous theorem. Since
$\leq$-relevance is \Dlog{3}-hard~\cite[Theorem~4.3.2]{eite-gott-95-a}, also
checking for solutions of  $NEXT\_SOL_\leq(\l H,M,T \r, \{A_1,\ldots,A_m\})$
containing a given $h \in H$ is \Dlog{3}-hard.~\qed

 %

 %

\section{Other Minimal Solution}

The implicit assumption in second-best solutions is that non-minimal solutions
are taken into account once all minimal ones have been considered. Indeed, the
definition of $NEXT\_SOL(\l H,M,T \r, \{A_1,\ldots,A_m\})$ includes all
solutions that are minimal once $A_1,\ldots,A_m$ are removed from
consideration. A different approach is to only allow minimal solutions. This is
different in that:

\begin{itemize}

\item second-best solutions are solutions that are minimal among the ones
different from the given ones;

\item other minimal solutions are solutions that are minimal and are not among
the given ones.

\end{itemize}

The difference is that the first definition allows non-minimal solutions if the
minimal ones are all among the given ones. The second definition does not. The
difference only concerns non-minimal solutions. Therefore, it disappears when
the void preorder $\void$ is considered, as no solution is non-minimal
according to it.

When using $\subseteq$ or $\leq$, the two definitions may lead to different
results, like in the problem:

\begin{eqnarray*}
H &=& \{s, r\} \\
M &=& \{t\} \\
T &=& \{s \rightarrow t\}
\end{eqnarray*}

The problem $\l H,M,T \r$ has two explanations: $\{s\}$ and $\{s,r\}$. Only the
first one is minimal in the two considered preorders; this is also intuitively
correct, as $r$ does not really contribute to entail $t$. However, the
second-best solutions include this non-minimal one: $NEXT\_SOL_\leq(\l H,M,T
\r, \{\{s\}\}) = \{\{s,r\}\}$. Such a possibility is excluded when considering
the other minimal solutions: no one exists apart from $\{s\}$.

When $\subseteq$ is used as the preorder, the complexity of checking another
minimal solution is the same as that for a second-best solution. This can be
proved as for the proof of Theorem~\ref{setc} with minimal changes: for
membership, sets $A \backslash \{h\}$ are checked even if they are in
$\{A_1,\ldots,A_m\}$; hardness is proved with the very same reduction, which
maps minimal solutions of the original problem into solutions of the new
problems that are both second-best solutions and other solutions.

Other best solutions are easier than second-best, if using $\leq$:
\Dp-complete. The following lemma shows how to relate the solutions of a
problem to the minimal solutions of another problem. This property will be used
to prove that we can reduce the problem of checking a solution to the problem
of checking another minimal solution.

\begin{lemma}
\label{equalizer}

Let $P = \l H,M,T \r$ be a problem of abduction, where $H=\{h_1,\ldots,h_n\}$.
Let $P'=\l H',M',T' \r$ be the problem defined as follows, where $C$, $D$, and
$E$ are sets of $n$ fresh variables each.

\begin{eqnarray*}
H' &=& C \cup D \\
M' &=& M \cup E \\
T' &=& T \cup
  \{ c_i \rightarrow h_i,
     c_i \rightarrow e_i,
     d_i \rightarrow e_i ~|~ 1 \leq i \leq n\}
\end{eqnarray*}

It holds:

\[
SOL_\leq(\l H',M',T' \r) =
\{
\{ c_i ~|~ h_i \in A \} \cup
\{ d_i ~|~ h_i \not\in A \} ~|~
A \in SOL(\l H,M,T \r)
\}
\]

\end{lemma}

\proof Intuitively, $e_i \in M'$ enforces either $c_i$ or $d_i$ to be in every
solution, and minimization excludes solutions containing both. Since every
$c_i$ entails $h_i$, $M$ is entailed only if the $c_i$'s correspond to the
original solutions. Since a solution not containing $c_i$ contains $d_i$, each
solution of $P$ is mapped into a minimal solution of $P'$.

Formally, the claim is proved in three steps: in the first, every solution of
$P$ is proved to be translatable into a solution of $P'$; in the second, every
solution of $P'$ can be translated back to a solution of $P$; in the third,
every minimal solution of $P'$ is shown to contain $d_i$ if and only if it does
not contain $c_i$. These three steps prove the claim.

\


Let $A$ be a solution of $P$, and $A'= \{ c_i \in C ~|~ h_i \in A \} \cup \{
d_i \in D ~|~ h_i \not\in A \}$. The first step of the proof is to show that
$A'$ is a solution of $P'$. Since $A \cup \{T\}$ is consistent, it has a model.
It can be extended to the new variables by setting $c_i$ to the same value of
$h_i$ and all $d_i$'s and $e_i$'s to true. This model satisfies $A$ and $T$,
and also all implications $c_i \rightarrow h_i$ because $c_i$ is true if and
only if $h_i$ is true, and $c_i \rightarrow e_i$ and $d_i \rightarrow e_i$
because $e_i$ is true. Therefore, $A' \cup \{T'\}$ is consistent.

Entailment $A' \cup \{T'\} \models M \cup E$ also holds. Since $A$ is a
solution of the original problem, $A \cup \{T\} \models M$ holds. Since $A'$
contains every $c_i$ such that $h_i \in A$, and $T'$ contains $c_i \rightarrow
h_i$, it follows that $A' \cup \{T'\} \models A$. As a result, $A' \cup \{T'\}
\models M$. Since $A'$ contains either $c_i$ or $d_i$ for every $i \in
\{1,\ldots,n\}$ by construction, and $T'$ contains $c_i \rightarrow e_i$ and
$d_i \rightarrow e_i$, it follows that $A' \cup \{T'\} \models E$. This proves
that $A'$ is a solution of $P'$.

\


The second step is to prove that every solution $A'$ of $P'$ can be translated
back to a solution of $P$. In particular, this holds with $A = \{ h_i ~|~ c_i
\in A'\}$. Consistency of $A \cup \{T\}$ is a consequence of the consistency of
$A' \cup \{T'\}$, since this formula contains $T$, $A' \cap C$ and $\{c_i
\rightarrow h_i\}$, the latter two implying $A$.

Entailment $A \cup \{T\} \models M$ is a consequence of $A' \cup \{T'\} \models
M'$ and $M \subseteq M'$, which imply $A' \cup \{T'\} \models M$. This holds if
and only if $A' \cup \{T',\neg m_i\}$ is inconsistent for every $m_i \in M$. In
this set, $e_i$ only occurs in $c_i \rightarrow e_i$ and $d_i \rightarrow e_i$,
unnegated in both. As a result, these two clauses can be removed without
affecting consistency. After this operation, if $d_i$ still occurs is in $A'$,
unnegated. It can therefore be removed. What remains is the following set,
which can be simplified by the usual methods:

\begin{eqnarray*}
\lefteqn{
(A' \cap C) \cup \{c_i \rightarrow h_i ~|~ 1 \leq i \leq n\} \cup \{T,\neg m_i\}
} \\
& \equiv &
(A' \cap C) \cup \{h_i ~|~ c_i \in A'\} \cup \{c_i \rightarrow h_i ~|~ c_i
\not\in A\} \cup \{T,\neg m_i\} \\
& \equiv &
(A' \cap C) \cup A \cup  \{c_i \rightarrow h_i ~|~ c_i \not\in A'\} \cup
\{T,\neg m_i\}
\end{eqnarray*}

Each $c_i$ occurs in a single clause: if $c_i \in A'$ then $c_i$ is only in $A'
\cap C$; if $c_i \not\in A'$ then it is only in $c_i \rightarrow h_i$. As a
result, all clauses containing $c_i$ can be removed without affecting
consistency, leading to $A \cup \{T,\neg m_i\}$. This proves that $A \cup \{T\}
\models m_i$. This holds for every $m_i \in M$; therefore, $A \cup \{T\}
\models M$.

\


The final part of the proof is to show that all minimal solutions contain
either $c_i$ or $d_i$ but not both. This claim can be divided in two: that no
solution lacks both $c_i$ and $d_i$ for some $i$, and that every solution that
contains both is not minimal.

Let $A'$ be a solution that contains neither $c_i$ nor $d_i$ for an arbitrary
index $i$. The set $A' \cup \{T',\neg e_i\}$ contains $c_i$ and $d_i$ only in
the clauses $c_i \rightarrow h_i$, $c_i \rightarrow e_i$ and $d_i \rightarrow
e_i$, negated in all. As a result, these clauses can be removed without
affecting consistency. The consequence of this deletion is that $e_i$ only
occurs negated, and can therefore be removed. What remains is a subet of $A'
\cup \{T'\}$, which is consistent because $A'$ is a solution. This proves that
$e_i$ is not entailed, contradicting the assumption that $A'$ is a solution.

Solutions of $P'$ may contain both $c_i$ and $d_i$ for some $i$. However, this
solution is not minimal, since $d_i$ can be removed from it. Let $A'$ be a
solution containing both $c_i$ and $d_i$. Since $A' \cup \{T'\}$ is consistent,
so is $A' \backslash \{d_i\} \cup \{T'\}$. Remains to prove that $A' \backslash
\{h\} \cup \{T'\} \models M'$, which amounts to the inconsistency of $A'
\backslash \{h\} \cup \{T', \neg \bigwedge M'\}$. Since $c_i \in A$, then $c_i
\rightarrow e_i$ implies $e_i$. As a result, $e_i$ can be added to the set, and
$d_i \rightarrow e_i$ removed. What remains is a formula that contains $d_i$
only unnegated, as part of $A'$. It can therefore be removed without affecting
inconsistency.~\qed

This lemma maps each solution of $P$ into a $\leq$-minimal solution of $P'$,
and viceversa. It therefore provides a reduction from the problem of
second-best solutions with the void preorder $\void$ to the problem of other
minimal solution with the cardinality preorder $\leq$.

\begin{theorem}

The problem of checking another minimal solution \wrt\  $\leq$ is \Dp-complete.

\end{theorem}

\proof Given $\{A_1,\ldots,A_m\}$ with $m \geq 1$, one can check whether $A$ is
another minimal solution by expressing $|A|=|A_1|$ as a propositional formula
$F$ using fresh variables. Then, the problem amounts to the satisfiability of
$A \cup \{T,F\}$ and the unsatisfiabity of $A \cup \{T,\neg \bigwedge M\}$.

Hardness follows the \Dp-hardness of the problem of verifying $A \in
NEXT\_SOL_\void(\l H,M,T \r,\{A_1,\ldots,A_m\})$. Indeed, Lemma~\ref{equalizer}
proves that solutions $A,A_1,\ldots,A_m$ of $\l H,M,T \r$ can be turned into
$\leq$-minimal solutions $A',A_1',\ldots,A_m'$ of $\l H',M',T' \r$. As a
result, $A$ is a solution of $\l H,M,T \r$ not in $\{A_1,\ldots,A_m\}$ if and
only if $A'$ is a minimal solution of $\l H',M',T' \r$ not in
$\{A_1',\ldots,A_m'\}$. Since the first problem is \Dp-hard, the second is
\Dp-hard as well.~\qed

 %

\section{Using Additional Information}

In the previous sections we have shown that the abduction problems remain
intractable even if we know a first solution. It seems that knowing a solution
does not help in reducing the computational complexity. In this section we
investigate whether during the search for the first solution, we could obtain
and store additional information (not just the solution) that allows for a
faster search for another solution. The complexity of such a problem can be
evaluated using compilability classes~\cite{cado-etal-02} and self
reductions~\cite{libe-01-jacm}.

In short, a problem has the same complexity with and without additional
information if the part of the problem instance the additional information
derives from can be ``moved'' to the rest of the instance; this is called a
compilability self reduction; more details are in
Section~\ref{complexity-compilability} and the cited articles. For abduction,
the additional information comes from $\l H,M,T \r$, the rest of the instance
is the subset $A \subseteq H$ to check.

The problems analyzed in the previous sections have the same complexity if $T$
is restricted to be a 3CNF: a set of clauses, each comprising exactly three
literals. Since $T$ is now a set, $\gamma \in T$ can be used to indicate that
the clause $\gamma$ is in $T$. Let $\var(T)$ be the set of all propositional
variables used by $T$, that is, the alphabet of $T$.

Given a set of variables $X$ (for example, $X=\var(T) \cup H \cup M$ in the
following proofs), $\Pi_X$ denotes the set of all possible clauses of three
literals over alphabet $X$. If $|X|=n$, the number of possible literals is
$2n$; this means that the number of possible clauses of three literals is less
than $2n \times 2n \times 2n=8 \times n^3$, a polynomial in $n$. The clauses of
$\Pi_X$ are considered enumerated, and called
$\gamma_1,\gamma_2,\gamma_3,\ldots$. Self reductions for problems of logics
usually employ this construction.

The first application of this concept is to the problem of verification with
the void preorder.

\begin{lemma}

If $P = \l H,M,T \r$ is a problem of abduction with $T$ in 3CNF and $A
\subseteq H$ let $P' = \l H',M',T' \r$ and $A'$ be defined as follows, where
$X=\var(T) \cup H \cup M$ (hence, $T \subseteq \Pi_X$) and $C$ is a set of
fresh variables in one-to-one correspondence with $\Pi_X$.

\begin{eqnarray*}
A' &=& A \cup \{ c_i ~|~ \gamma_i \in T \} \\
H' &=& H \cup C \\
M' &=& M \\
T' &=& \{ c_i \rightarrow \gamma_i ~|~ \gamma_i \in \Pi_X \}
\end{eqnarray*}

It holds:

\[
A \in SOL(\l H,M,T \r) \mbox{ iff }
A' \in SOL(\l H',M',T' \r)
\]

\end{lemma}

\proof The first part of the proof is that $A \cup T$ is consistent if and only
if $A' \cup T'$ is consistent. Since $\{c_i, c_i \rightarrow \gamma_i\}$ is
equivalent to $\{c_i, \gamma_i\}$, it holds:

\begin{eqnarray*}
A' \cup T'
&\equiv&
A \cup \{c_i ~|~ \gamma_i \in T \} \cup
\{c_i \rightarrow \gamma_i ~|~ \gamma_i \in \Pi_X \} 			\\
&\equiv&
A \cup \{c_i ~|~ \gamma_i \in T \} \cup
\{c_i \rightarrow \gamma_i ~|~ \gamma_i \in T \} \cup
\{c_i \rightarrow \gamma_i ~|~ \gamma_i \in \Pi_X \backslash T \}	\\
&\equiv&
A \cup \{c_i ~|~ \gamma_i \in T \} \cup
\{\gamma_i ~|~ \gamma_i \in T \} \cup
\{c_i \rightarrow \gamma_i ~|~ \gamma_i \in \Pi_X \backslash T \}	\\
&\equiv&
A \cup \{c_i ~|~ \gamma_i \in T \} \cup T \cup
\{ c_i \rightarrow \gamma_i ~|~ \gamma_i \in \Pi_X \backslash T \}
\end{eqnarray*}

In this formula, each $c_i$ appears once, either in $\{c_i ~|~ \ldots\}$ or in
$\{c_i \rightarrow \gamma_i ~|~ \ldots\}$. As a result, these clauses can be
removed without affecting satisfiability. The result is $A \cup \{T\}$, proving
that this set and $A' \cup \{T'\}$ are equisatisfiable.

The second part of the proof is that $A \cup T \cup \{\neg \bigwedge M\}$ is
consistent if and only if $A' \cup T' \cup \{\neg \bigwedge M\}$ is consistent.
Thanks to the above chain of equivalences, $A' \cup T'$ can be rewritten as $A
\cup \{c_i ~|~ \gamma_i \in T \} \cup \{T\} \cup \{ c_i \rightarrow \gamma_i
~|~ \gamma_i \in \Pi_X \backslash T \}$. Therefore:

\[
A' \cup T' \cup \{\neg \bigwedge M\} \equiv
A \cup \{c_i ~|~ \gamma_i \in T \} \cup T \cup
\{ c_i \rightarrow \gamma_i ~|~ \gamma_i \in \Pi_{\var(T)} \backslash T \}
\cup \{\neg \bigwedge M\}
\]

Again, each $c_i$ only occur once in this formula. Therefore, all clauses
containing it can be removed, leading to the equisatisfiable formula $A \cup T
\cup \{\neg \bigwedge M\}$. Therefore, $A' \cup T' \models M$ if and only if
$A \cup T \models M$.~\qed

This lemma provides a self-reduction for the problem of solution checking for
the empty preorder. In order to derive a proof of compilability hardness from
it, the three functions of classification, representativeness and extensions
are needed.

In this section, all abduction problems are assumed to be built over an
alphabet $H_n \cup M_n \cup X_n$ for some $n$, where:

\begin{eqnarray*}
H &=& \{h_1,\ldots,h_n\} 				\\
M &=& \{m_1,\ldots,m_n\}				\\
X &=& \{x_1,\ldots,x_n\}
\end{eqnarray*}

This is not a restriction: if the variables are not these ones, they can be
renamed; if $|H|<|M|$ new variables can be added to $H$; if $|\var(T)
\backslash H \backslash M|<|H|$ new variables can be added to $T$; for $M$, the
new variables are also added to $T$.

The classification, representative and extension functions are defined over
pairs $\l A, \l H,M,T \r\r$ where $\l H,M,T \r$ is a problem of abduction and
$A \subseteq H$ a candidate solution for it. The class of the pair $I=\l A, \l
H,M,T \r \r$ is its number of assumptions, why coincide with its number of
manifestations and the number of other variables in the instance.

\[
Class(I)=|H|
\]

The representative instance of the class $n$ has $n$ variables of each type:

\[
Repr(m)= \l \emptyset, \l H_n, M_n, \Pi_{H_n \cup M_n \cup X_n} \r \r
\]

The extension function is obtained by adding new variables. If $Class(I)=n$ and
$m>n$ then:

\begin{eqnarray*}
\lefteqn{Ext(\l A,\l H,M,T \r\r,m) = \l A,\l H',M',T' \r\r \mbox{ where:}} \\
H' &=& H \cup \{h_{n+1}, \ldots, h_m\} \\
M' &=& M \cup \{m_{n+1}, \ldots, m_m\} \\
T' &=& T \cup \{m_{n+1}, \ldots, m_m\} \cup \{x_{n+1}, \ldots, x_m\}
\end{eqnarray*}

This instance has $m$ assumptions, meaning that $Class(\l A,\l H,M,T \r\r)=m$,
as required to the extension function. The second requirement is that of
equivalence: $A$ is a solution of $\l H,M,T \r$ if and only if $A$ is a
solution of $\l H',M',T' \r$. This holds because in $A \cup T'$ and $A \cup T'
\cup \{\neg \bigwedge M\}$ the new variables $h_{n+1},\ldots,h_m$ only occur
once (in $A$), the new variables $x_{n+1},\ldots,x_m$ only once (in $T$), and
the new variables $m_{n+1},\ldots,m_m$ in $T$ and $M$ but unnegated in both.
All these new variables can therefore be removed without affecting consistency.

This proves the existence of the classification, representative and extension
function for the problem of second-best solution verification. Since solutions
are not changed by the extension function, these can be used with all of the
considered preorders: void, set-based and cardinality-based.

The following results require problems of abductions to be restricted to the
case where the formula $T$ is in 3CNF. Lemma~\ref{empty-hard} and
Theorem~\ref{cardinality-checking} instead employ reductions that produce come
clauses that have more than three literals. In particular, the first turns $G$
into $\neg G[X'/X] \rightarrow m$ and the second introduces $\bigwedge S
\rightarrow \bigwedge M'$. Both can be turned into clauses, but in general with
more than three literals. The following lemma shows how to turn a formula in
3CNF without altering the abductive solutions.

\begin{lemma}
\label{syntactic}

If $l_1$ and $l_2$ are two literals, $C$ a clause and $x$ a fresh variable,
then
{} $SOL(\l H,M,T \wedge (l_1 \vee l_2 \vee C) \r) =
{}  SOL(\l H,M,T \wedge (l_1 \vee l_2 \vee x) \wedge (\neg x \vee C) \r)$.

\end{lemma}

\proof Every model $M$ of $T \wedge (l_1 \vee l_2 \vee C)$ satisfies either
$l_1 \vee l_2$ or $C$. A model of
{} $(l_1 \vee l_2 \vee x) \wedge (\neg x \vee C)$
can be constructed by setting $x$ to false in the first case and to false in
the second. In the other way around, if $M$ is a model of
{} $(l_1 \vee l_2 \vee x) \wedge (\neg x \vee C)$
then it assigns $x$ to either true or flase. In the first case $M$ satisfies
$C$, in the second $l_1 \vee l_2$.

This not only proves that the two formulae are equisatisfiable, but that they
have the same models apart from the value of $x$. Since $x \not\in H$ and $x
\not\in M$, it follows that
{} $B \cup \{T \wedge (l_1 \vee l_2 \vee C)\}$
and
{} $B \cup \{(l_1 \vee l_2 \vee x) \wedge (\neg x \vee C)\}$
are also equisatisfiable for every $B \subseteq H \cup \{\neg m ~|~ m \in M\}$.
Since the abductive solutions are defined in terms of the satisfiability of $T$
with a subset of $H$ with possibly the negation of an element of $M$, the claim
is proved.~\qed

A simple iteration of this lemma to all clauses of $T$ made of more than three
literals proves that the problems of abduction are unchanged by the restriction
to clauses of three literals.

 %

\begin{theorem}

The problem of deciding whether $A \subseteq H$ is in $SOL(\l H,M,T \r)$ is
\nuc{\Dp}-complete.

\end{theorem}

\proof Membership follows from that in \Dp, which was proved in a previous
section, and the fact that every compilability class \nuc{C} contains the
relative complexity class C~\cite{cado-etal-02}.

Let $\l A,\l H_n,M_n,T \r\r$ be a pair of class $n$. By definition, the
representative element of the class $n$ is a pair $\l A',\l H',M',T' \r\r$ in
the same class $n$. The class being the same implies that $H'=H_n$, $M'=M_n$
and $\var(T') \backslash H' \backslash M'=\var(T') \backslash H' \backslash
M'$. In other words, $\l H',M',T'\r$ has the same hypotheses $H_n$,
manifestations $M_n$ and other variables $X_n$ of $\l H,M,T \r$.

A reduction satisfies representative equivalence if and only if $\l A,\l
H_n,M_n,T \r\r$ and $\l A,\l H',M',T'\r\r$ are translated into equivalent
instances. In both pairs the candidate solution is $A$, but in the second the
problem of abduction $\l H,M,T \r$ is replaced by the one of the representative
instance $\l H',M',T \r$. The reduction of the previous lemma translates $\l
A,\l H_n,M_n,T \r\r$ and $\l A,\l H',M',T'\r\r$ into the same pair $\l A,\l
H'',M'',T''\r\r$, since $A$ is translated into $A$ and the problem of abduction
into one that depends only on its variables; since $\l H,M,T \r$ and $\l
H',M',T \r$ have the same variables, they are translated into the same problem.
The results of translation are therefore the same instance, which means that it
is a self reduction. Since the problem of checking whether $A$ is a solution of
$\l H,M,T \r$ is \Dp-hard even in the restriction of clauses of three literals
thanks to Lemma~\ref{syntactic} and has the required classification,
representativeness and extension functions, it is also \nuc{\Dp}-hard.~\qed

The lemma provides a reduction from solutions to solutions, but cannot be used
with $\subseteq$ and $\leq$, as $A$ may not be minimal because of another
explanation $A'$ that does not contain a $c_i \in A$. The point is that $c_i
\in A$ indicates the presence of $\gamma_i \in T$, and should therefore not be
included in the minimization.

The problem is solved using a construction similar to that of
Lemma~\ref{equalizer}: for each $c_i$ introduce an hypothesis $d_i$ and a
manifestation $e_i$, and the clauses $c_i \rightarrow e_i$ and $d_i \rightarrow
e_i$ in $T$. This way, the variables $c_i$ are not considered in the
minimization.

\begin{lemma}

Given $P = \l H,M,T \r$ and $A \subseteq H$, construct $P'$ and $A'$ as
follows.

\begin{eqnarray*}
A' &=& A \cup
       \{ c_i ~|~ \gamma_i \in T \} \cup \{ d_i ~|~ \gamma_i \not\in T \} \\
H' &=& H \cup C \cup D \\
M' &=& M \cup E \\
T' &=& \{ c_i \rightarrow e_i ~|~ c_i \in C \} \cup
       \{ d_i \rightarrow e_i ~|~ d_i \in D \} \cup
       \{ c_i \rightarrow \gamma_i ~|~ \gamma_i \in \Pi_{\var(T)} \}
\end{eqnarray*}

The sets $C$, $D$, and $E$ are sets new variables in one-to-one correspondence
with $\Pi_{\var(T)}$, where $\var(T)$ is the set of propositional variables of
$T$. It holds:

\[
A \in SOL_\subseteq(\l H,M,T \r) \mbox{ iff }
A' \in SOL_\subseteq(\l H',M',T' \r)
\]

\end{lemma}

The proof is omitted because of its similarity with that of
Lemma~\ref{equalizer}. The instance that results from this transformation can
be further modified as explained above to make the number of assumptions,
manifestations and other variables to be the same.

The following theorem shows that the case of set-containment is not different
from the case of the empty preorder, in the sense that compiling $\l H,M,T \r$
does not lower complexity.

\begin{theorem}

The problem of checking solutions using $\subseteq$ is \nucP{2} complete.

\end{theorem}

\proof The problem is in \P{2}; therefore, it is also in \nucP{2}. Hardness is
proved by the reduction in the previous lemma: since $\l H',M',T' \r$ only
depends on the class of $\l A,\l H,M,T \r\r$, this translation satisfies the
condition of representative equivalence. The classification, representative,
and extension functions are the ones shown before. Since the problem is
\P{2}-hard this proves that it is also \nucP{2}-hard ~\qed

The problem with $\leq$ is \nucDp-complete. Indeed, from $\l H,M,T \r$ one can
calculate the size of its minimal solutions, and then use this number to
determine whether a set of hypotheses is a minimal solution. The previous lemma
provides a proof of hardness for the same class, in the same way as in the
previous theorem. The proof is omitted because of its similarity with the
previous one.

\begin{theorem}

Checking whether a solution is minimal \wrt\  $\leq$ is \nucDp-complete.

\end{theorem}

 %

\section{Conclusions}

In this article, we have investigated the problem of finding a solution to a
given abduction problem when some solutions have already been found. The
results show that the analyzed problems are computationally intractable, but
this does not rule out the possibility of tackling them. It only suggests the
most appropriate tools to use. Polynomial problems are best attacked using
deterministic polynomial algorithms, while problems in NP can be solved using
reduction to the propositional satisfiability problem (SAT) and then passed to
a state of the art SAT solver (for example, one of the contestants in the SAT
competition \url{http://www.satcompetition.org/}). Problems in higher classes
of the polynomial hierarchy (such as all the problems shown in the paper) can
be solved by a reduction to the Quantified Boolean Formulae problem (QBF) and
the use of QBF solvers (\url{http://qbf.satisfiability.org/gallery/}). Problems
higher up in the polynomial hierarchy are more complex to solve, but, by
identifying the precise complexity, we can better take advantage of the
solvers.

There are some open questions and some possible future directions of work. It
makes sense to establish the complexity of finding a $k$-th best solution, at
least in the case of ordering based on cardinality. This can be seen as a
variant of the problems studied in this article where the given solutions are
not known.

Another question left open by this article is to find a reduction from the
problem of second-best solutions to simple abductions that preserve the
explanations. What is needed is the opposite of Lemma~\ref{adding}, which shows
how to add a given explanation to an abduction problem: a reduction that
eliminates some given solutions from an abduction problem while leaving the
other ones unchanged.

 %

\bibliographystyle{plain}

\end{document}